# Probing topological protection using a designer surface plasmon structure

Fei Gao[1#], Zhen Gao[1#], Xihang Shi[1], Zhaoju Yang[1], Xiao Lin[1,2], Hongyi Xu[1], John D. Joannopoulos[3], Marin Soljačić[3], Hongsheng Chen[2,3,*], Ling Lu[3], Yidong Chong[1,4,*], Baile Zhang[1,4,*]

[1]Division of Physics and Applied Physics, School of Physical and Mathematical Sciences, Nanyang Technological University, Singapore 637371, Singapore.

[2]State Key Laboratory of Modern Optical Instrumentation, Zhejiang University, Hangzhou 310027, China.

[3]Department of Physics, Massachusetts Institute of Technology, Cambridge, Massachusetts 02139, USA.

[4]Centre for Disruptive Photonic Technologies, Nanyang Technological University, Singapore 637371, Singapore.

#These two authors contributed equally to this work.

*Authors to whom correspondence should be addressed; E-mail: hansomchen@zju.edu.cn (H. Chen); yidong@ntu.edu.sg (Y. Chong); blzhang@ntu.edu.sg (B. Zhang)

**Topological photonic states, inspired by robust chiral edge states in topological insulators, have recently been demonstrated in a few photonic systems, including an array of coupled on-chip ring resonators at communication wavelengths. However, the intrinsic difference between electrons and photons determines that the 'topological protection' in time-reversal-invariant photonic systems does not share the same robustness as its counterpart in electronic topological insulators. Here, in a designer surface plasmon platform consisting of tunable metallic sub-wavelength structures, we construct photonic topological edge states and probe their robustness against a variety of defect classes, including some common time-reversal-invariant photonic defects that can break the topological protection, but do not exist in electronic topological insulators. This is also the first experimental realization of anomalous Floquet topological edge states, whose topological phase cannot be predicted by the usual Chern number topological invariants.**

A topological insulator (TI) is a material that is electrically insulating in the bulk, but conducts along the surface via a family of 'topological edge states', whose existence is guaranteed by the topological incompatibility between the TI's electronic band structure and the vacuum[1,2]. Recently, a novel field of 'topological photonics' has emerged, seeking to exploit the phenomenon of topological photonic states using classical electromagnetic (EM) waves[3,4]. These states are not only promising for defect-resistant EM wave-guiding applications, but also provide a unique platform for fundamental studies of the physics of topological phases that are not easily available in the condensed-matter context.

The first demonstration of topological photonic states was in a microwave-scale two-dimensional (2D) photonic crystal containing magneto-optic elements biased by an external magnetic field to break time-reversal symmetry[5,6]. Two time-reversal-invariant designs, operating at optical frequencies, followed. Firstly, Rechtsman *et al.*[7] demonstrated an array of three-dimensional (3D) coupled helical waveguides in which the paraxial propagation along the third dimension mapped formally to a periodically-driven, or 'Floquet,' 2D TI[8]. Secondly, Hafezi *et al.*[9-11] realized a time-reversal-invariant photonic TI in the form of an on-chip lattice of coupled optical ring resonators, engineered to simulate a uniform magnetic field in the quantum Hall effect. Many other designs have also been proposed recently.[12-14] However, the intrinsic difference between electrons and photons determines that the 'topological protection' in such time-reversal-invariant photonic systems[7,9-13,15] does not share the same robustness as its counterpart in electronic topological insulators, and the limits require further experimental study. Furthermore, the topological phases in previous photonic systems[7,9,12] all have existing condensed-matter realizations which can already provide experimental platforms[7,10,13] to study them. The usefulness of topological photonics to understanding fundamental topological physics can be demonstrated with the explicit construction of a novel topological phase that still lacks a condensed-matter realization.

Here, we report on the implementation of a topological designer surface plasmon structure operating in the microwave regime. Designer surface plasmons (also called 'spoof surface-plasmons')[16-20] are electromagnetic modes analogous to the familiar plasmon modes which occur in metallic surfaces and resonators at infrared and optical frequencies; these modes, however, appear at much lower frequencies and are supported by the presence of periodic sub-wavelength corrugations in the underlying metal structures. They hold considerable promise in microwave- to infrared-frequency device applications, due to the ease with which their properties can be fine-tuned by altering the underlying structural parameters. This highly-tunable platform allows us to probe the robustness of topological edge states under a wide variety of different defect conditions. Furthermore, this system is a realization of an anomalous Floquet TI, which is a topological phase that has not yet been realized in a condensed-matter setting.

Using the designer surface plasmon platform, we study the performance of the photonic TI in the presence of specific defects. Firstly, we show that the topological edge states are indeed immune to backscattering from a variety of barriers that do not reverse the propagation of edge states, including removal of entire unit cells. However, the path detour of propagation around defects can deteriorate the transmission of topological edge states, because of the intrinsic propagation loss of designer surface plasmons. Next, we construct two kinds of common photonic defects, without counterparts in electronic TIs, to break the topological protection of these edge states. First of all, for electronic edge states under topological protection in TIs, a default law is particle number conservation[1,2]. In contrast, photons can be easily annihilated or created by loss or gain in photonic systems[21]. Although loss is already present in above defect studies, here we construct an extreme case, a strongly dissipative defect that can completely annihilate photonic topological states without backscattering. This kind of annihilation defect has not been observed in electronic topological insulators[1,2]. Second, it is well-known that a 'magnetic,' or time-reversal-breaking, defect in TIs can flip the spin of electrons and cause backscattering for topological edge states. In periodically driven Floquet TIs, a defect cannot flip the time harmonic modulation on the whole structure, and thus cannot reverse the propagation of topological edge states. On the other hand, photonic pseudo spins/modulations effectively realized with wave circulation can be flipped by some common time-reversal-invariant defects. We demonstrate that the photonic TI is not topologically robust against such defects; in a practical photonic TI device, therefore, such defects need to be explicitly suppressed.

Finally, it is worth noting that even though our study is based on an anomalous Floquet photonic TI[23], other photonic TIs which have been demonstrated experimentally[7,10,13] have the same features and limitations. In particular, both Hafezi *et al.*'s pseudo spin approach and Rechtsman *et al.*'s Floquet modulation approach are effectively realized with wave circulation that can be coupled to the opposite circulation. The coupled ring resonator lattice studied by Hafezi *et al.*[9-11] relies on having decoupled photonic pseudo spins. Likewise, the coupled helical waveguide arrays studied by Rechtsman *et al.*[7] rely on propagation down one direction along the waveguide axis which determines an effective circulation direction in the reduced 2D lattice as a time harmonic modulation; backscattering along the axial direction, which would induce the opposite circulation, is ignored. Our systematic experimental tests of topological protection including the possibility of flipping the pseudo spin or Floquet modulation are therefore applicable to these systems as well.

**Results**

**Implementing topological designer surface plasmon structure.** The designer surface plasmon structure is shown in Fig. 1a. It consists of closely-spaced sub-wavelength metallic rods, placed on a flat metallic surface in an arrangement similar to the design of Hafezi *et al.*[9-11] (but with a significant conceptual difference in the topological phase, to be discussed in the next section). Large rings, called 'lattice rings', are set in a square lattice, and each pair of adjacent lattice rings is connected by a smaller 'coupling ring'. Designer surface plasmon waves can circulate clockwise or counter-clockwise in the lattice rings, serving as 'pseudo spins' in the photonic pseudo spin approach as demonstrated in previous designs[9-11]. For the moment, we consider the typical situation where the two circulations do not couple to each other. Modes of the chosen circulation can be excited via U-shaped input/output waveguides at the corners of the lattice (Fig. 1b). The field pattern is recorded by a near-field probe scanning above the metal rods, connected to a microwave network analyzer. (See Materials and Methods for details.)

**Mapping to anomalous Floquet topological insulator phase**. A significant departure from the design of Hafezi *et al.*,[9-11] where the coupling rings were assigned different geometries for constructing an incommensurate 'magnetic vector potential' to simulate the quantum Hall effect, is that the coupling rings in the present lattice have identical geometries. Therefore, the lattice is entirely commensurate, and cannot be mapped to a quantum Hall system. Nonetheless, its band structure is topologically non-trivial.[21-23] The lattice is described by a network model[21-25] which can be formally mapped onto a Floquet lattice, with the phase delay $\phi$ (as marked in Fig. 1a) in each quarter of a lattice ring playing the role of a Floquet quasi-energy[21-23]. When the coupling between adjacent lattice rings is increased beyond a critical value ($\theta = 0.25\pi$ in the parameterization of Ref. 21 and Ref. 23), the lattice undergoes a topological transition from a topologically trivial quasi-energy band structure to a topologically non-trivial one with robust topological edge states (Fig. 2c; details to be discussed later).

Before we elaborate on the theoretical modelling, we emphasize that, as the most unusual feature of this topologically non-trivial phase, the Chern numbers are zero for all bands, for each (decoupled) circulation[23]. Normally, in spin-decoupled condensed-matter and photonic systems, the net number of topological edge states in each band gap is equal to the sum of the Chern numbers in all bands below the gap ('bulk-edge correspondence')[1-5,7,12-13,26]. However, 'anomalous Floquet TI' phases are an exception[27-29] for the following reason. Because Floquet quasi-energies, unlike ordinary energies, are angle variables[27-29], Floquet band structures are thus not bounded below by a 'lowest band'. In other words, the infinite number of bands below the gap makes it impossible to apply the usual Chern-number-based bulk-edge correspondence. Although the corresponding non-vanishing topological invariant of the anomalous Floquet TI phase has recently been experimentally confirmed[30], an extended 2D lattice supporting this exotic topological phase has not been demonstrated. Our designer surface plasmon structure thus serves as the first explicit realization of an anomalous Floquet topological phase.

We start the theoretical modelling with a 2D illustration of this periodic lattice as depicted in Fig. 2a. Each lattice ring acts as a waveguide, constraining EM waves to propagate along the ring. When the coupling through the coupling ring (scaled down in Fig. 2a) has negligible internal backward scattering, the mode-hopping from a lattice ring to neighboring lattice rings conserves the circulation. To determine the bandstructure, we consider one unit cell of the lattice, shown schematically in Fig. 2b. In each lattice ring, we define a complex four-vector whose components are the input amplitudes $|a_n\rangle = [a_{1n}, a_{2n}, a_{3n}, a_{4n}]$, and another four-vector containing the output amplitudes $|b_n\rangle = [b_{1n}, b_{2n}, b_{3n}, b_{4n}]$. These input and output amplitudes are related by $|a_n\rangle = e^{-i\phi}|b_n\rangle$, where $\phi$ is the phase delay along a quarter lattice ring. For Bloch modes, which satisfy $|a_n\rangle = |a_K\rangle e^{iK \cdot r_n}$ and $|b_n\rangle = |b_K\rangle e^{iK \cdot r_n}$, the inter-cell scattering can be described by $S(K)|b_K\rangle = |a_K\rangle$, where $S(K)$ is a unitary scattering matrix derived from the couplings between the lattice and coupling rings; it is periodic in $K$ with the periodicity of the Brillouin zone. (See Supplementary Information for details.) We thus obtain the governing scattering matrix equation

$$S(K)|b_K\rangle = e^{-i\phi}|b_K\rangle$$

We can regard the eigenvectors $|b_n\rangle$ in the above equation as Bloch eigenstates; then $\phi(K)$ plays the role of a band energy, except for the fact that it is an angle variable ($\phi \equiv \phi + 2\pi$). We refer to $\phi$ as the 'quasienergy.'

The Bloch modes of a periodic network are equivalent to the Floquet modes of a periodically driven lattice. Suppose we have a lattice, of the same spatial dimensions as our network, with a Hamiltonian $H_K(t)$ that is periodic in time with period $T$. Then the Floquet state with state vector $|b_n\rangle$ and Floquet quasienergy $\phi(K)/T$ obeys exactly the governing scattering matrix equation, provided $S(K)$ is the time-evolution operator over one period:

$$S(K) = T\exp\left[-i\int_0^T dt H_K(t)\right]$$

where $T$ is the time-ordering operator. This relationship between network models and Floquet lattices was pointed out in Ref. 23. One can regard $S(K)$ as a discrete time-evolution operator acting on a particle which is initially localized at one point in the network (say the midpoint of a quarter lattice ring as the link); over one time period, the particle moves along the link, tunnels instantaneously across a node, and moves midway along a neighbouring link.

Using the above formalism, we have calculated the quasi-energy band structure of a semi-infinite strip with 50 lattices in *y* direction and periodic in *x* direction; the results are shown in Fig. 2c. (See Supplementary Information for additional details.) By tuning the effective inter-ring coupling strength $\theta$, we can achieve a topological phase transition. For weak couplings $\theta < 0.25\pi$, the band structure is gapped; the gaps close at a critical value $\theta = 0.25\pi$, and for strong couplings $\theta > 0.25\pi$ the gaps re-open with topologically protected edge states, while the Chern numbers of all bands are still zero, as verified numerically.

**Simulation of designer surface plasmons.** We implement designer surface plasmons by arranging periodic metallic rods on a metal surface. The dispersion of the designer surface plasmons can be tuned by changing the rod heights. Fig. 3a shows a 1D periodic array of metallic rods with periodicity $p = 5.0$ mm on a flat metallic surface. Each metallic rod has radius $r = 1.25$ mm and the same height *h* (*h* can be varied). The background is free space. We simulate the dispersion relation of the guided designer plasmon waves, for different values of *h*; the results are shown in Fig. 3b. The mode profile of the electric field, for $h = 5.0$ mm, is shown in Fig. 3c. We observe that the fields are tightly confined around the rods; other choices of *h* give similar mode profiles. This strong waveguiding makes it feasible to probe the field distribution using a near-field scanning measurement.

Only part of the quasi-energy band structure shown in Fig. 2c is accessible, because the designer surface plasmons propagate only within a narrow frequency band as in Fig. 3b. As it is unfeasible to simulate the band structure of the full 3D structure *ab initio*, we adopt an alternative approach based on repetitive frequency scanning, which shortens the computation time; see Materials and Methods for details. Fig. 4a shows the simulated band structure for a semi-infinite strip which has 5 lattice rings in the *y* direction, and is infinite in the *x* direction (choosing the circulation where the modes run clockwise along the lattice rings). These results reveal a gap between 11.1 GHz and 11.7 GHz, spanned by unidirectional states localized to opposite edges of the strip. Note that within such a narrow bandwidth, the inter-ring coupling strength $\theta$ is mainly determined by the spacing *g* between lattice and coupling rings. We estimate that for the current setting with *g*=5.0 mm, the coupling strength is $\theta = 0.41\pi \pm 0.03\pi$ within the band gap of interest (see Supplementary Information). According to calculation from the network model in Fig. 2, this band gap is topologically nontrivial.

**Demonstration of topological protection and its robustness.** We now experimentally study the topological edge state in a finite 5×5 lattice. First, we apply a monopole source to the bulk, at a mid-gap (11.3 GHz) frequency; this produces a mode localized in the vicinity of the source (Fig. 4b), verifying that the bulk is insulating. Next, we excite the structure via one of the U-shaped input/output waveguides at 11.3 GHz. This produces a mode which propagates along the edge (Fig. 4c), including around one corner of the lattice. No obvious reflection is observed in experiment. The transmission reaches -12.94 dB at the output as shown in Fig. 4d. High transmission within a frequency range around 11.3 GHz corresponds extremely well with the band gap predicted in Fig. 4a.

The transmission drop of 12.94 dB arises from the propagation loss over 9 lattice constants. We thus estimate that the propagation loss per lattice constant is about 1.44 dB at 11.3 GHz. Finally, in order to verify that the observed topological edge state is a consequence of the bulk structure in the 2D lattice, we remove all the rings except those along the bottom and left boundaries, which form a one-dimensional (1D) chain (Fig. 4e2-3). The defect ring is left in place along this chain. The mode is now strongly reflected by the defect ring (Fig. 4e1), and the transmission at the output reaches the noise level (Fig. 4d).

To probe the robustness of the topological edge state, we introduce a defect by altering one of the lattice rings along the edge, decreasing the height of its rods from 5.0 mm to 3.5 mm (Fig. 4f2). As can be seen in the dispersions in Fig. 3b, this decrease of height forms a sharp momentum mismatch when the edge state intends to go through this defect ring. As a result of topological protection, the resultant edge state circumvents the defect ring, and continues propagating along the modified edge (Fig. 4f1). However, as can be seen in Fig. 4d, the path detour of propagation leads to the transmission drop from -12.94 dB to -19.86 dB at the output. This roughly 7 dB drop is consistent with the propagation loss over extra 5 lattice constants as a result of the path detour.

The robustness of topological protection can be further demonstrated by varying the height of the defect ring, as analogues of defects with different 'potential barriers' in electronic materials. The first defect variation is achieved by fully removing a lattice ring (or equivalently, reducing its height to zero) and its surrounding coupling rings (Fig. 5a3), being similar to the defect in Ref. 10. Because of the zero probability for the waves to couple to this defect, it corresponds to an 'infinite potential barrier'.

Simulation (Fig. 5a1) and experiment (Fig. 5a2) show that the edge mode circumvents the defect of missing rings without touching. The transmission at the output reads -16.04 dB, as summarized in Table 1. Compared to the case with no defect, the path detour leads to a longer propagation length with 2 more lattice constants (shorter than the case of 3.5 mm-tall defect ring), which drops the transmission for about 3 dB.

The second defect variation consists of a lattice ring of metallic rods with 4.3 mm height (Fig. 5b3). Since the dispersion in the modified lattice ring is close to a regular lattice ring, this defect can be considered as a low potential barrier (substantially weaker than the similar defect in Fig. 4f that can be treated as a medium potential barrier). The low potential barrier allows part of the mode to directly tunnel through this defect, while the remainder still circumvents the defect, as can be seen clearly in simulation (Fig. 5b1) and experiment (Fig. 5b2). Therefore, the transmission at the output reads -17.4 dB (see the summarized measurement in Table 1), at the level between cases of 3.5 mm-tall defect ring and no defect.

**Demonstration of breaking topological protection.** The above defects, similar to previous demonstrations in topological photonics, do not break topological protection. The first defect that we will demonstrate to break the topological protection is a backscattering-immune strongly dissipative defect, which is realized by gradually decreasing heights of the rods in the defect ring (inset of Fig. 6a). As simulated in Fig. 6a, the scattering occurring at this defect is able to fully dissipate, or 'annihilate,' the edge states, while backscattering is suppressed due to the adiabatic momentum change on the rods with gradually decreasing heights. The measured field pattern in Fig. 6b confirms this dissipating phenomenon. The stronger radiation in Fig. 6b than in Fig. 6a is because in experiment the monopole probe of finite length collects fields in a finite range of heights, while in simulation only fields at a single height are captured. The transmission at the output reaches noise level.

We then demonstrate the second defect, which breaks the topological protection by reversing the propagation of edge states. This propagation reversal of edge states corresponds to spin flip in the photonic pseudo spin approach[9-11] or the modulation reversal in the Floquet modulation approach[7]. Hereafter we adopt the 'spin' flip picture to describe this process for convenience. First, we implement a partial 'spin' flip defect by approaching a metallic block to the lattice ring without touching (Fig. 7a3). This defect can mix the two pseudo 'spins'. Similar defects such as a semi-transparent scatterer[9] have been proposed previously, but have not been demonstrated. For the purpose of imaging the propagation of 'spin' flipped modes, we deliberately decrease the height of input waveguide from 5.0 mm to 4.3 mm, such that the coupling of 'spin' flipped mode to the input waveguide is weak. We estimate that the coupling loss from the input waveguide to the structure is about 7 dB. (See in Supplementary Information the estimation on coupling efficiency). Both the simulated (Fig. 7a1) and measured (Fig. 7a2) field patterns reveal partial mixing between the two pseudo 'spins,' with a portion of the mode with clockwise circulation continuing to propagate to the left and eventually exiting from the lower leg of U-shape waveguide at the upper left corner, while the remainder is reflected to the right with counter-clockwise circulation, eventually exiting from the upper leg of the U-shape waveguide. Because signals reaching the output waveguide are too weak to observe, we measure the transmission/reflection at the location marked by a white horizontal line in Fig. 7a2, where the transmitted and reflected waves have the same propagation length. The measured transmission/reflection at -20.25 dB/-18.36 dB shows that the ratio between the transmitted 'spin' down state and the reflected 'spin' up state is about 1:1.54.

Second, we construct a defect capable of flipping the 'spin' completely by inserting the metallic block between two metallic rods in a lattice ring (Fig. 7b3). Simulation in Fig. 7b1 shows a complete conversion from the clockwise circulation, or pseudo 'spin-down,' to the counter-clockwise circulation, or pseudo 'spin-up,' by the defect at the bottom edge of the structure. The edge state only exits from the lower leg of U-shape waveguide at the upper left corner. The measured field pattern (Fig. 7b2) matches well with the simulated results. The measured reflection of -16.29 dB at the location on the marked horizontal line in Fig. 7b2 is consistent with the transmission in Fig. 4c without defect after considering the coupling loss at the input and the reduced propagation length in experiment.

**Demonstration of a topologically trivial phase.** For completeness, we finally demonstrate the behaviour of the designer surface plasmon structure in the topologically trivial phase. As shown in Fig. 2c, the network band structure is topologically trivial when the coupling between lattice rings is sufficiently weak. To accomplish this, we increase the inter-ring separation $g$ from 5 mm to 7.5 mm. We first simulate the band diagram as shown in Fig. 8a. It can be seen that between the two bulk bands, there is no edge state in the bandgap which spans from 11.35 GHz to 11.5 GHz. When the structure is excited at 11.3 GHz (in a bulk band), an extended field pattern is observed in both simulation (Fig. 8b) and experiment (Fig. 8d). When the excitation is tuned to 11.45 GHz (in the band gap), the field patterns in both simulation (Fig. 8c) and experiment (Fig. 8e) show a mode that is localized in the vicinity of excitation and does not propagate. The retrieved coupling strength within this frequency band of interest shows weaker coupling strength than the critical value of topological phase transition (See Supplementary Information).

**Discussion**

The above results, as summarized in Table 1, demonstrate the topological protection and its robustness against various defects on a time-reversal-invariant photonic platform of designer surface plasmons. When facing the defects analogous to low (defect no. 1), medium (defect no. 2), and infinite (defect no. 3) potential barriers, the topological edge states can circumvent these defects without back-scattering. However, the path detour still deteriorates the transmission because of propagation loss over elongated path. On the other hand, the strongly dissipative defect (defect no. 4) and the metallic block (defects no. 5 and 6) can break the topological protection for completely dissipating the edge states, and inducing reflection, respectively. The reflection corresponds to spin flip in Hafezi *et al.*'s pseudo spin approach[9-11] and modulation reversal in Rechtsman *et al.*'s Floquet modulation approach[7]. Furthermore, the anomalous Floquet topological phase constructed in this study cannot be predicted by the usual Chern number topological invariants. Our demonstration makes the first step to explicitly construct this exotic topological phase in a physical photonic structure. In view of the tunability of the designer surface plasmon structure, intriguing avenues to explore in the future include the possibility of topologically protected mode amplification[21], which can be achieved by integrating microwave amplifiers, and topological many-body physics by incorporating non-linear components.

**Materials and Methods**

The 5×5 lattice of lattice rings contains a total of 3,320 metal rods, each having diameter 2.5 mm and height 5.0 mm, standing on a flat aluminum plate which is 1 m × 1 m in size and 5.0 mm thick. Each lattice ring consists of 56 rods arranged in a circle of radius of $R_1$=44.56 mm. Adjacent lattice rings are coupled through coupling rings, each consisting of 48 rods arranged in a circle of radius $R_2$=38.20 mm.

The excitation source is a single-mode cable-to-waveguide adaptor with a rectangular port. Half of the port is covered with aluminum foil to increase the cutoff frequency of the waveguide port. Exciting the U-shape waveguide at different legs (input 1 or 2) can excite different circulations of surface EM wave in lattice rings. The field pattern of $E_z$ component is recorded with a 3 mm-in-length monopole probe which scans in the *xy* plane, 1 mm above the top of the metal rods.

The commercial software CST Microwave Studio is used for numerical simulation. In band structure simulation, a supercell of the designer surface plasmon structure that consists of 5 lattice rings in *y* direction and 1 lattice ring in *x* direction is adopted. The periodic boundary condition is imposed with a phase shift in *x* direction. An external dipole source close to the top ring is used to excite the network. By scanning the frequency, each dip of the receiving spectrum (widely termed '$S_{11}$' parameter) of this dipole source should correspond to a mode of this network for a specific phase shift in *x* direction. By further scanning the phase shift through the whole Brillouin zone, the complete band structure can be obtained. Circulation of the edge states can be monitored with dynamic field distributions.

**Acknowledgements**

This work was sponsored by Nanyang Technological University for NAP Start-up Grant, Singapore Ministry of Education under Grant No. MOE2015-T2-1-070 and MOE2011-T3-1-005, and the Singapore National Research Foundation under Grant No. NRFF2012-02. The work at MIT was supported by the U. S. Army Research

Laboratory and the U. S. Army Research Office through the Institute for Soldier Nanotechnologies, under contract number W911NF-13-D-0001, in part by the MRSEC Program of the NSF under Award No. DMR-1419807, and in part by the MIT S3TEC EFRC of DOE under Grant No. DE-SC0001299. The work at ZJU was supported by the NNSFC (Grants Nos. 61322501, 61574127, and 61275183), the National Top-Notch Young Professionals Program, the NCET-12-0489, and the FRFCU-2014XZZX003-24. We are grateful to M. Rechtsman for helpful comments.

**Author Contributions**

All authors contributed extensively to this work. F. G. and Z. G. designed structures and performed measurements. F. G. implemented transfer matrix method. Z. G. and H. X. performed simulations. X. S., Z. Y., and X. L. involved in scanner setup and assembling sample. F. G., Z. G., J. D. J., M. S, H. C., L. L., Y. C. and B. Z. analyzed data, discussed and interpreted detailed results, and prepared the manuscript with input from all authors. H. C., L. L., Y. C. and B. Z. supervised the project.

**Competing Financial Interests statement**

The authors declare no competing financial interests.

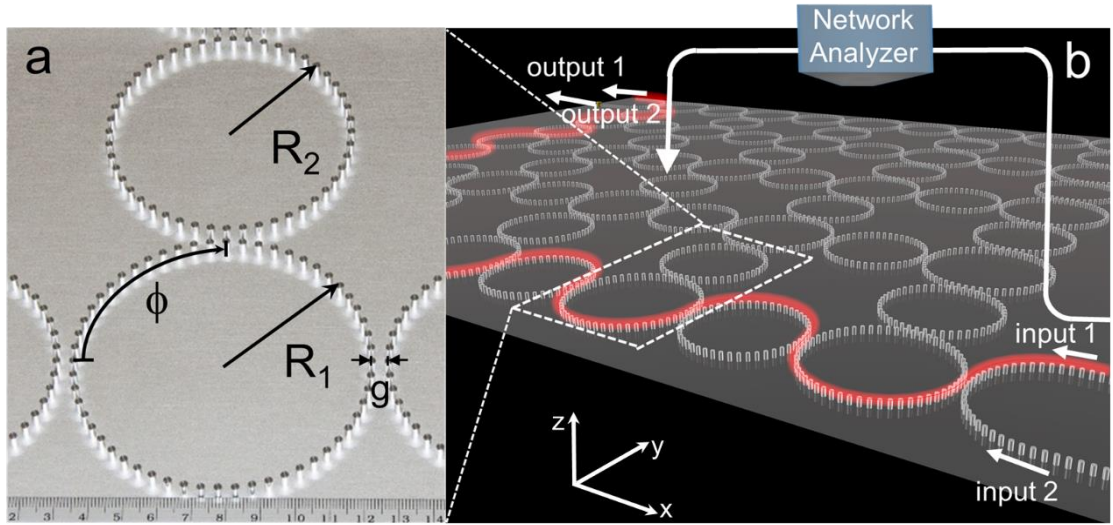

**Figure 1 | Construction of a topological designer surface plasmon structure. a,** Photo of metallic rods with diameter 2.5 mm and height 5.0 mm distributed with center-center distance 5.0 mm on a flat metallic surface. A lattice ring formed by 56 metallic rods is with radius $R_1$ = 44.56 mm. A coupling ring formed by 48 metallic rods is with radius $R_2$ =38.2 mm. The ring-ring distance is $g$ = 5.0 mm. $\phi$ denotes the phase delay of electromagnetic waves along a quarter of a lattice ring. **b,** Schematic of a 5×5 lattice in experiment. A network analyzer records the field pattern by scanning a near-field probe above the metallic rods. The red meandering curve above the structure represents edge states.

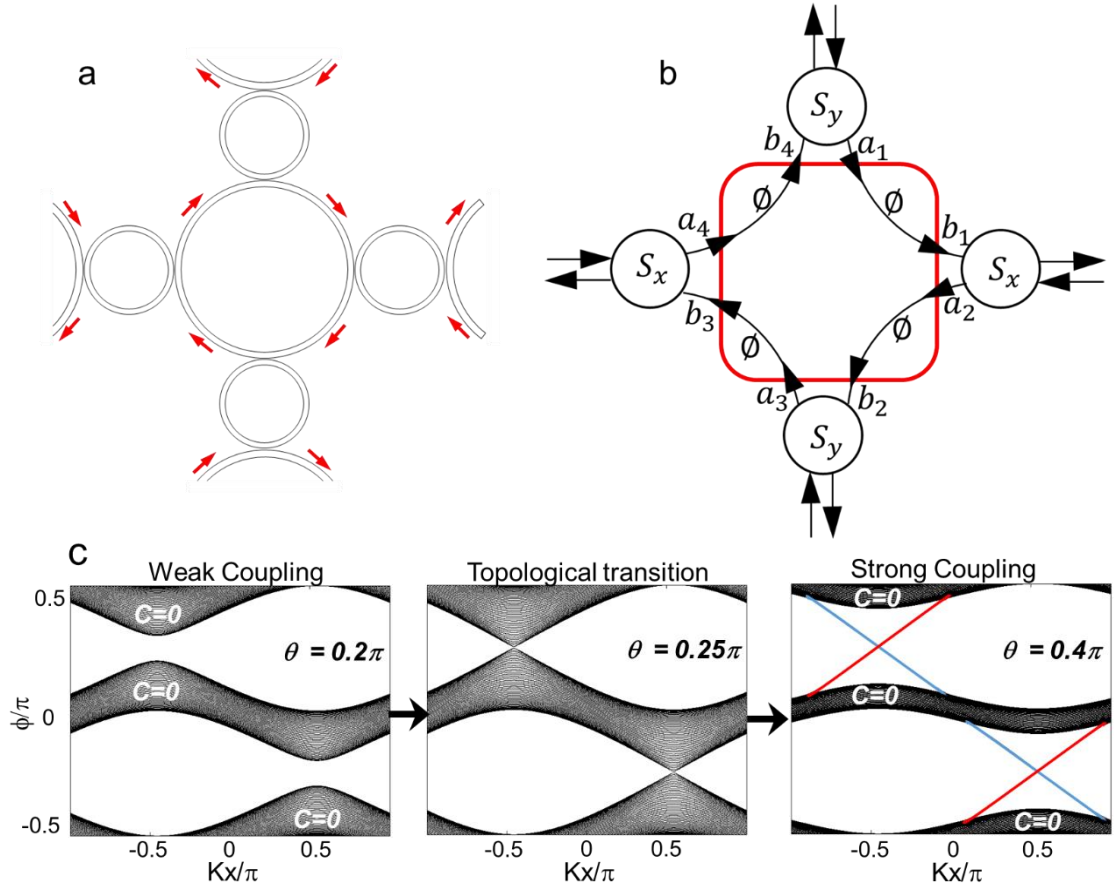

**Figure 2 | Network model description of the topological structure and its topological transition. a,** Schematic of a unit cell in a two-dimensional lattice of coupled ring waveguides. **b,** The equivalent periodic network. Within the unit cell, we define a surface (red rectangle) which is penetrated by input amplitudes $|a\rangle$ and output amplitudes $|b\rangle$, related by $|b\rangle = e^{-i\phi}|a\rangle$. These amplitudes also scatter with those of neighbouring cells, with coupling matrices $S_x$ and $S_y$. **c,** Topological transition as the inter-ring coupling strength θ is tuned from weak to strong. Before and after the transition, all bands have zero Chern number C=0. Red and blue lines denote edge states confined to the upper and lower edges of the strip respectively.

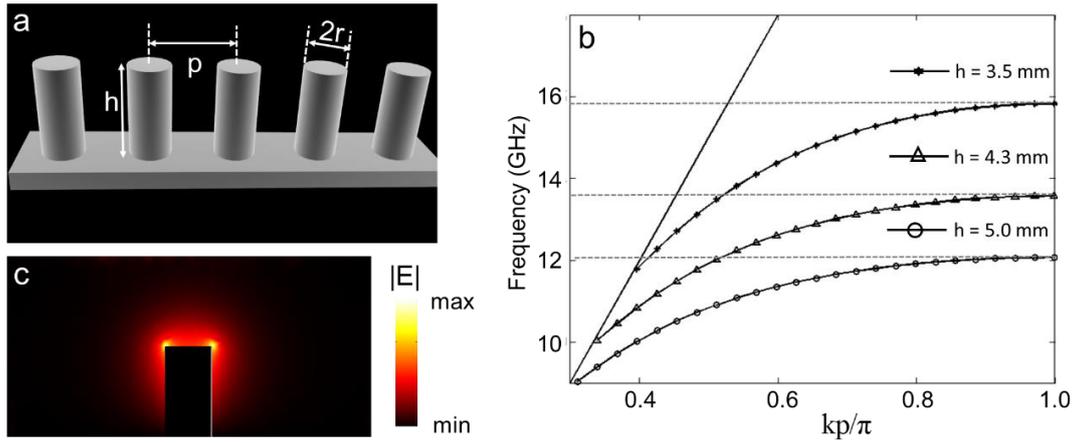

**Figure 3 | Dispersion and field profile of designer surface plasmons. a,** Schematic of an array of metallic rods on a flat metallic surface. **b,** Dispersions of designer surface plasmons with heights of rods $h = 3.5$ mm (star curve), $h = 4.3$ mm (triangle curve), and $h = 5.0$ mm (circle curve), respectively. The black solid line is the light line. **c,** Mode profile of |E| field in the cross section of a metallic rod that is perpendicular to the waveguiding direction ($h = 5.0$ mm).

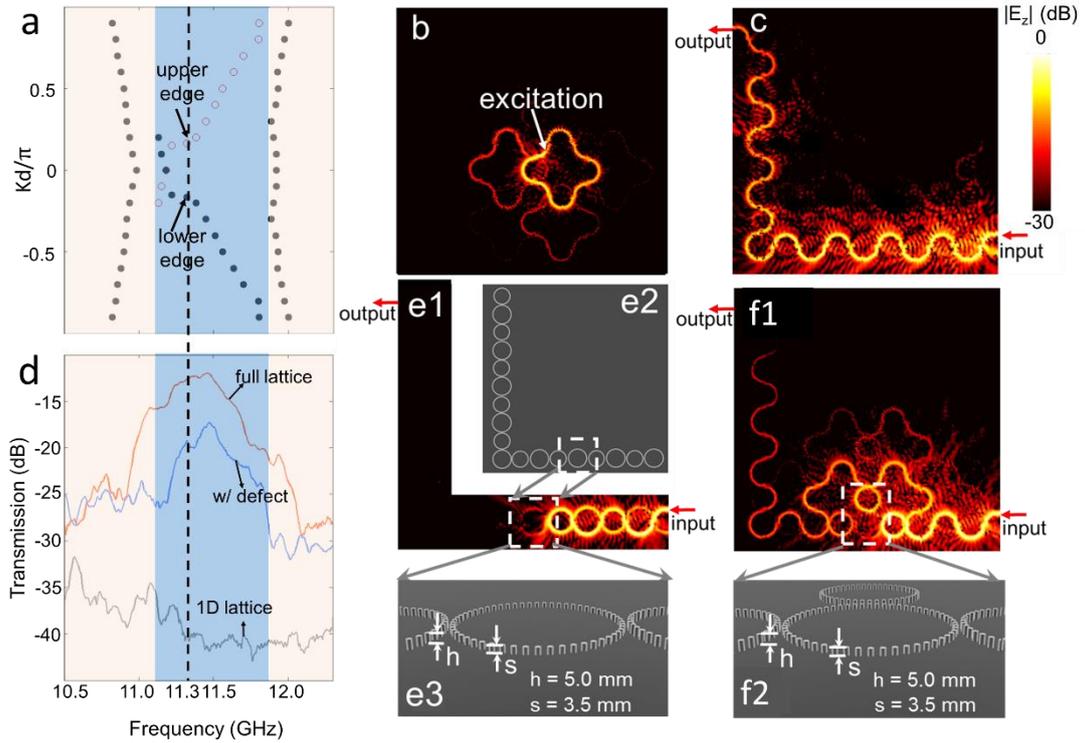

**Figure 4 | Demonstration of topological edge state. a,** Simulated band diagram of the narrow-band designer surface plasmon structure. **b,** Observed field pattern when the excitation is inside the bulk at frequency 11.3 GHz. **c,** Observed edge state at frequency 11.3 GHz. **d,** Transmission spectra of configurations in **c, d,** and **f**. **e1,** The defect (shown in **e3**) causes strong reflection for a 1D lattice (**e2**). **f1,** The edge state circumvents a defect lattice (shown in **f2**) consisting of metallic rods with height $s = 3.5$ mm. All other rods have the height $h = 5.0$ mm.

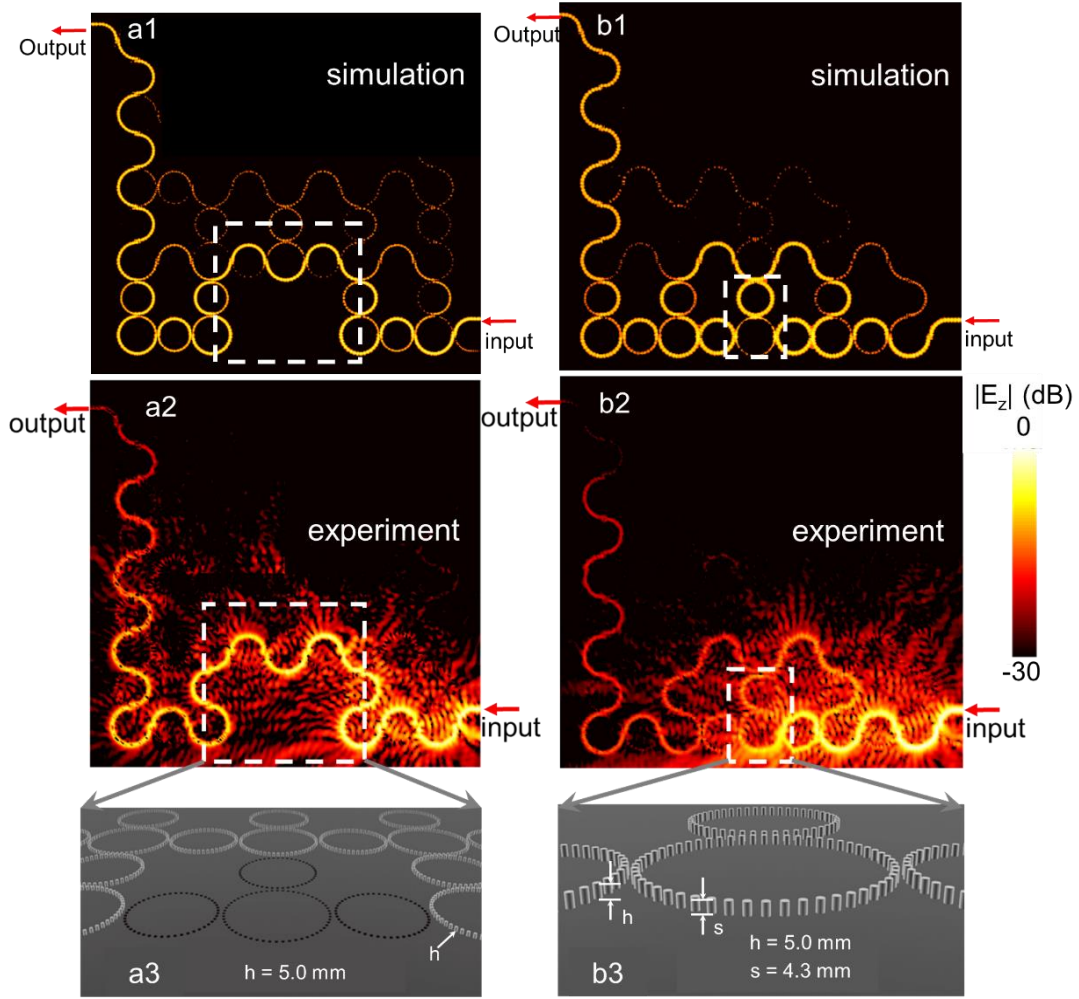

**Figure 5 | Demonstration of the robustness of topological protection against defect variations. a1-a3,** Simulated (**a1**) and observed (**a2**) field pattern when removing a lattice ring and its surrounding coupling rings (shown in **a3**). **b1-b3,** Simulated (**b1**) and observed (**b2**) field pattern when the height of metallic rods of a lattice ring is decreased to $s = 4.3$ mm (shown in **b3**). All other rods maintain their height $h = 5$ mm. Metal is modelled as a perfect electric conductor in simulation.

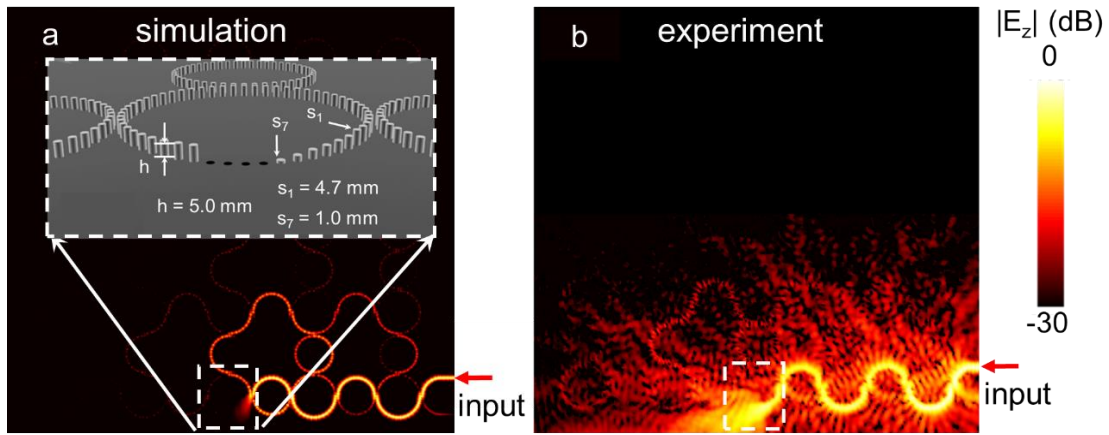

**Figure 6 | Demonstration of backscattering-immune strongly dissipative defect. a,** Simulated field pattern when metallic rods of a lattice ring gradually decrease their heights to zero (shown in the **inset;** from $s_1$ to $s_7$, $s_1$ = 4.7 mm, $s_2$ = 4.1 mm, $s_3$ = 3.5 mm, $s_4$ = 2.9 mm, $s_5$ = 2.3 mm, $s_6$ = 1.7 mm, and $s_7$ = 1.0 mm). Metal is modelled as a perfect electric conductor in simulation. **b,** Observed field pattern corresponding to simulation in **a**.

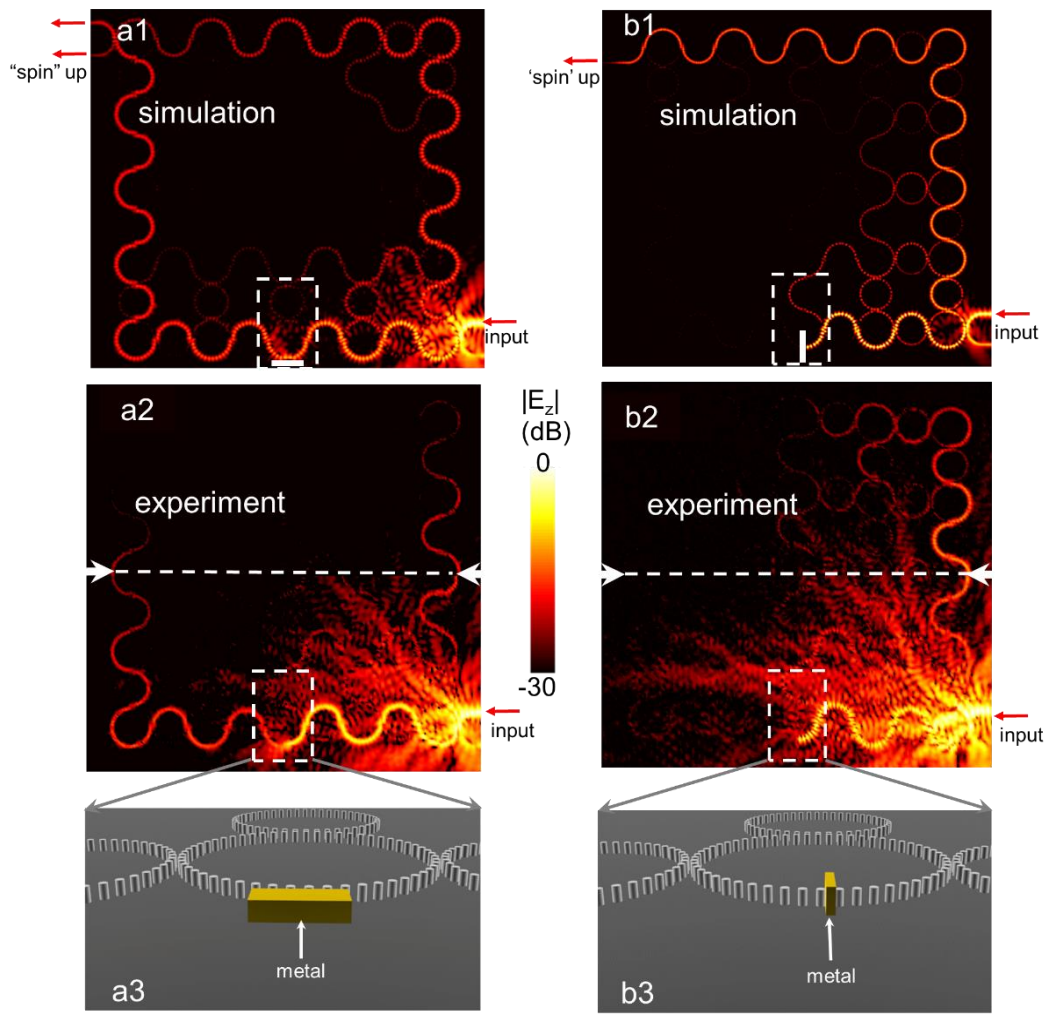

**Figure 7 | Demonstration of time-reversal-invariant 'spin' flipping defect. a1-a3,** Simulated (**a1**) and observed (**a2**) field pattern when a metallic block (shown in **a3**) is located near a lattice ring without touching. **b1-b3,** Simulated (**b1**) and observed (**b2**) field pattern when the metallic block (shown in **b3**) is inserted in a lattice ring. Metal is modelled as a perfect electric conductor in simulation.

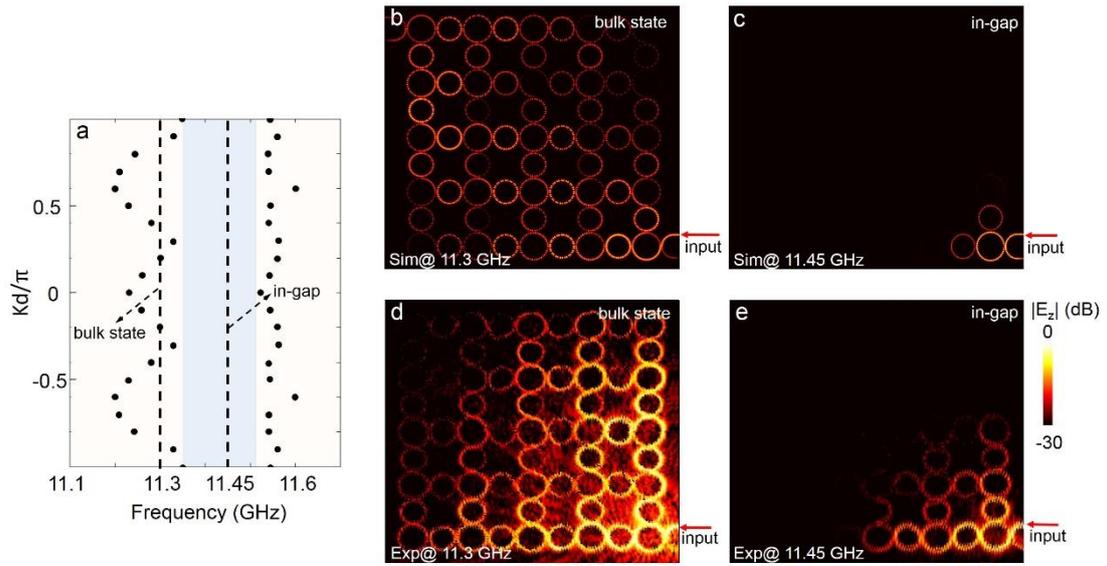

**Figure 8 | Demonstration of a trivial insulator phase. a,** Simulated band diagram of the topologically trivial designer surface plasmon structure. **b,** Simulated field pattern of bulk state at 11.3 GHz. **c,** Simulated field pattern at 11.45 GHz in the band gap. **d,** Observed field pattern at 11.3 GHz in the bulk band. **e,** Observed field pattern at 11.45 GHz in the topologically trivial band gap. Metal is modelled as a perfect electric conductor in simulation.

| Defect No. | Figure | Defect type | Physical meaning | Transmission | Reflection | Comment |
|---|---|---|---|---|---|---|
| 0 | Fig. 4c | No defect | Perfect lattice | -12.94 dB | <-35 dB | Measuring propagation loss |
| 1 | Fig. 5b2 | Rod height decreased to 4.3 mm | Low potential barrier | -17.4 dB | <-35 dB | Path detour and tunneling |
| 2 | Fig. 4f1 | Rod height decreased to 3.5 mm | Medium potential barrier | -19.86 dB | <-35 dB | Path detour |
| 3 | Fig. 5a2 | Rings removed | Infinite potential barrier | -16.04 dB | <-35 dB | Shorter path detour |
| 4 | Fig. 6b | Rod heights gradually decreased | Dissipation | <-35 dB | <-35 dB | >35 dB dissipation at defect |
| 5 | Fig. 7a2 | Metallic block approached | Partial flip of spin/modulation | -20.25 dB | -18.36 dB | ~7 dB coupling loss at input |
| 6 | Fig. 7b2 | Metallic block inserted | Complete flip of spin/modulation | <-35 dB | -16.29 dB | ~7 dB coupling loss at input |

**Table 1 | Summary of testing results on topological protection against defects.**